\documentclass[a4paper,useAMS,usenatbib]{mnras}

\pdfoutput=1

\usepackage{mathptmx}
\usepackage[T1]{fontenc}
\usepackage{ae,aecompl}

\usepackage[dvips]{graphicx}
\usepackage{amsmath}
\usepackage{amsfonts}
\usepackage{amssymb}
\usepackage{comment}
\usepackage[normalem]{ulem}
\usepackage[dvipsnames]{xcolor}
\usepackage[%
        ]{hyperref}
        
\hypersetup{
	colorlinks=true,	% false: boxed links; true: colored links
	urlcolor=MidnightBlue,		% color of external links
	pdfpagelabels=true,
	hypertexnames=true,
	plainpages=false,
	naturalnames=true,
	pdftitle={A hardCORE model for the minimum core size of solid exoplanets},    % title
	pdfauthor={Suissa et al.},     % author
	linkcolor=WildStrawberry,          % color of internal links (change box color with linkbordercolor)
	citecolor=ForestGreen,        % color of links to bibliography
}

\newcommand{\Kepler}{{\it{Kepler}}}

\newcommand{\httplink}{\href{https://github.com/gsuissa/hardCORE}{this URL}}

\newcommand{\hardcore}{{\tt HARDCORE}}
\newcommand{\crf}{\mathrm{CRF}}
\newcommand{\crfmin}{\crf_{\mathrm{min}}}
\newcommand{\crfmax}{\crf_{\mathrm{max}}}
\newcommand{\crfmarg}{\crf_{\mathrm{marg}}}

\newcommand{\orob} {R}
\newcommand{\re}{R_{\oplus}}

\newcommand{\omob} {M}
\newcommand{\me}{M_{\oplus}}

\title[A \hardcore\ model for constraining an exoplanet's core size]{A \hardcore\ model for constraining an exoplanet's core size}
\author[Suissa et al.]{Gabrielle Suissa$^{1}$\thanks{E-mail:
\href{mailto:ge2205@columbia.edu}{ge2205@columbia.edu}}, Jingjing Chen$^{1}$ \& David Kipping$^{1}$\\
$^{1}$Dept. of Astronomy, Columbia University, 550 W 120th Street, New York NY 10027}

\date{Accepted 2018 February 5. Received 2017 December 29; in original form 2017 September 5}

\pubyear{2017}

\begin{document}
\label{firstpage}
\pagerange{\pageref{firstpage}--\pageref{lastpage}}
\maketitle

\begin{abstract}
The interior structure of an exoplanet is hidden from direct view yet likely
plays a crucial role in influencing the habitability of Earth analogs.
Inferences of the interior structure are impeded by a fundamental degeneracy
that exists between any model comprising of more than two
layers and observations constraining just two bulk parameters: mass and radius.
In this work, we show that although the inverse problem is indeed degenerate,
there exists two boundary conditions that enables one to infer the minimum and maximum core radius fraction, $\crfmin$ \& $\crfmax$.
These hold true even for planets with light volatile envelopes, but
require the planet to be fully differentiated and that layers denser than iron
are forbidden. With both bounds in hand, a marginal CRF can also be inferred by
sampling inbetween. After validating on the Earth, we apply our method to
Kepler-36b and measure $\crfmin=(0.50\pm0.07)$, $\crfmax=(0.78\pm0.02)$
and $\crfmarg=(0.64\pm0.11)$, broadly consistent with the Earth's true CRF value of
0.55. We apply our method to a suite of hypothetical measurements of synthetic
planets to serve as a sensitivity analysis. We find that $\crfmin$ \&
$\crfmax$ have recovered uncertainties proportional to the relative error on
the planetary density, but $\crfmarg$ saturates to between 0.03 to 0.16 once
$(\Delta\rho/\rho)$ drops below 1-2\%. This implies that mass and radius alone
cannot provide any better constraints on internal composition once bulk density
constraints hit around a percent, providing a clear target for observers.
\end{abstract}

\begin{keywords}
planets and satellites: interiors --- planets and satellites: terrestrial planets --- planets and satellites: composition
\end{keywords}

\section{Introduction}
\label{sec:intro}

Despite recent strides in our ability to characterize exoplanets
\citep{kaltenegger:2017}, knowledge regarding the internal structure of distant
worlds remains almost entirely lacking \citep{spiegel:2014,baraffe:2014}.
Unlike the search for exoplanetary atmospheres \citep{burrows:2014}, moons
\citep{kipping:2014} or tomography \citep{mctier:2017}, our remote observations
do not have direct access to that which we seek to infer - the planet's
interior. The habitability of an Earth-like planet, in particular via the
likelihood of plate tectonics, is likely strongly influenced by the internal
structure \citep{noak:2014} and thus the community is strongly motivated
to infer what lies beneath, as part of the broader goal of understanding our own
planet's uniqueness.

In general, the only information we have about an exoplanet which is directly
affected by internal structure is the bulk mass and radius of the
planet\footnote{Quantities such as bulk density and surface gravity are of
course derivative of mass and radius}. Aside from this, we highlight that there
are some special cases where additional information about the planetary
interior can become available. For example, \citet{kaltenegger:2010} argue that
volcanism and planetary outgassing could be detectable using atmospheric
characterization techniques. Certain dynamical configurations of planetary
systems, such as tidal fixed points \citep{batygin:2009} for example, can also
enable inference of the planetary tidal properties, which in turn constrains
internal composition (see also \citealt{kramm:2012}). Finally, direct
measurements of oblateness may also provide constraints on internal structure
\citep{seager:2002,carter:2010,zhu:2014}.

Whilst there is some hope of identifying outgassing of exoplanets, providing
clues to the mantle composition \citep{kaltenegger:2010}, and measuring
tidal dissipation constants in special cases \citep{batygin:2009}, full 
structure inference will likely be limited to indirect methods based on
theoretical models. In this approach, one takes the basic observables we do
have access to, in particular planetary mass and radius, and compares them to
theoretical models in an effort to find families of compatible solutions. Since
theoretical models depend on more than just two parameters, accounting for
factors such as chemical composition \citep{valencia:2006,seager:2007},
ultraviolet environment \citep{lopez:2013,batygin:2013} and age
\citep{fortney:2011}, the problem is degenerate, in a general sense.

Since we do not have direct access to the interior of exoplanets, their
interior structure is generally modelled by assuming several key chemical
constituents. In the case of solid exoplanets, extrapolation from the Solar
System implies that they should be comprised of three primary chemical
ingredients, namely water, H$_2$O, enstatite, MgSiO$_3$, and iron, Fe
\citep{valencia:2006}. If we assume the planet is not young and has thus
become fully differentiated, the equations of state of these three layers can
be solved to provide theoretical estimates of the mass and radius of solid
bodies \citep{zeng:2013}. A fourth layer describing a light volatile
envelope can be placed on top to capture the behavior of mini-Neptunes,
where the light envelope is assumed to have negligible relative mass and
thus only affects the bulk radius and not the mass \citep{kipping:2013,
lopez:2014,wolfgang:2016}.

These four constituents can be combined in multiple ways to re-create the same
mass and radius. Even in the absence of a volatile envelope this
degeneracy persists, leading to the common use of ternary diagrams to
illustrate their symplectic yet degenerate loci of solutions (e.g. see
\citealt{seager:2007}). This degeneracy is a major barrier to inferring
unique solutions for planetary interiors, leading authors to either switch out
to simpler and non-degenerate two-layer models (e.g. \citealt{zeng:2016}) or
adding a chemical proxy from the parent star (e.g. \citealt{dorn:2017}) to
break the degeneracy. Whilst these are both certainly promising avenues for
tackling interior inference, in this work we focus on a third approach based
on boundary conditions.

The possibility of exploiting boundary conditions was first highlighted in
\citet{kipping:2013}, where the authors focused on the concept of
``minimum atmospheric height''. The method works by first predicting the
maximum allowed radius of a planet without any extended envelope given
its measured mass. This atmosphere-less planet is typically assumed to be
a pure water/icy body, for which detailed models are widely available. If
the observed radius exceeds this maximum limit, then some finite volume
of atmosphere must sit on top of the planetary interior, and the difference
in radii represents the ``minimum atmospheric height'' (MAH). The approach
therefore formally describes a key boundary condition of a general four-layer exoplanet.

In this work, we explore the other extreme, asking the question under what
conditions would an observed mass and radius definitively prove some
finite iron-core must exist, and what is the minimum radius fraction that
the core must comprise? Going further, we argue that the maximum core radius
fraction is another boundary condition in the problem and thus can be
derived to provide a complete bounding box of a planet's core size in a
general four-layer model framework.

We introduce the concept of the minimum core size in Section~\ref{sec:minimum},
as well as our fast parametric model to interpolate the \citet{zeng:2013} grid
models. Section~\ref{sec:inverting} discusses our approach to inverting the
relation to solve for core radius fraction directly, as well as the
much more straightforward method for inferring the maximum core size. In
Section~\ref{sec:applications}, we demonstrate the approach on both synthetic
and real exoplanets, with special attention to sensitivity. Finally, we discuss
the anticipated value of this work, as well its limitations, in
Section~\ref{sec:discussion}.

\section{Boundary Conditions on the Core}
\label{sec:minimum}

\subsection{Outline and Assumptions}
\label{sub:BC}

Exoplanets are generally expected to display a diverse range of physical
characteristics, owing to their presumably distinct formation mechanisms,
chemical environments and histories, as three examples. Even if the
mass and radius of an exoplanet were known to infinite precision, and the body
was known to be definitively solid, these two observed parameters are
insufficient to provide a unique solution for the relative fractions of
water, silicate and iron typically assumed to represent the major
constituents of solid planets \citep{kipping:2013}. In other words, mass and
radius alone cannot confidently reveal an exoplanet's CRF or CMF (core radius
fraction or core mass fraction, respectively).

\begin{figure}
\begin{center}
\includegraphics[width=8.4cm,angle=0,clip=true]{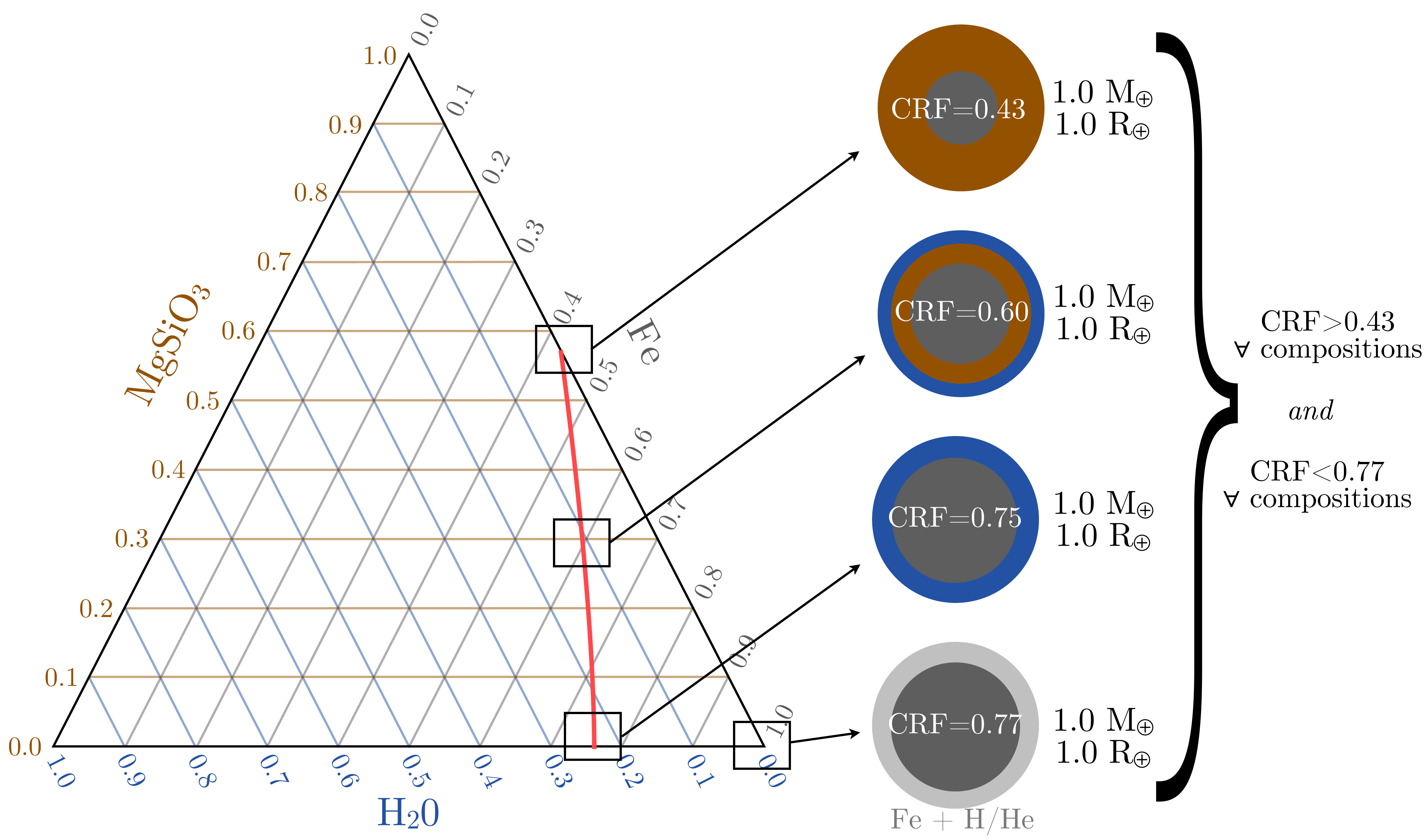}
\caption{
Ternary diagram of a three-layer interior structure model for a solid planet.
All points along the red line lead to a $M=1$\,$M_{\odot}$ and
$R=1$\,$R_{\odot}$ planet. Although indistinguishable from each other with
current observations, all points satisfy having a core-radius fraction
exceeding 43\%, a boundary condition we exploit in this work. The
largest iron core size allowed is depicted by the lowest sphere, where
the volatile envelope contributes negligible mass.
}
\label{fig:degeneracy}
\end{center}
\end{figure}

As a concrete example, a planet composed of water and iron can have the same
mass and radius as an iron-silicon planet, and thus have very different CRFs
(as illustrated in Figure~\ref{fig:degeneracy}).

As touched on in Section~\ref{sec:intro}, four-layer theoretical models of
solid planets are degenerate for a single mass-radius observation.
However, across the suite of loci able to serve as viable solutions, there
exists a boundary condition when the composition is pure silicate and iron. At
this point, the CRF takes the smallest value out of all possible models, since
the second-layer (the mantle) is now as heavy as it can be, being pure
silicate (the second densest material). Therefore, for any given mass-radius
pair, we can solve for the corresponding CRF of a pure silicate-iron model
(which is not a degenerate problem) and define that this CRF must equal the
minimum CRF, $\crfmin$, allowed across all models.

Similarly, another boundary condition we can exploit is to consider the maximum
allowed core size. As depicted in Figure~\ref{fig:degeneracy}, this occurs when
all of the mass is located within a pure iron core, padded by a second layer
of a light volatile envelope. The core can't possibly exceed this fractional
size else the mass would be incompatible with the observed value.

Note that although we refer to CRF rather than CMF here and in what follows,
once armed with a CRF, the mass and radius, it is easy to convert back to CMF.
Note too that here and throughout in what follows, we refer to the CRF strictly in
terms of an iron core. Although technically we acknowledge that a water-silicate
body could be described as having a finite sized silicate core, that core would
not qualify as being a ``core'' in this work.

Before continuing, we highlight some key assumptions of our model, for the sake
of transparency:

\begin{itemize}
\item[{\tiny$\blacksquare$}] The planet is fully differentiated and is not recently formed.
\item[{\tiny$\blacksquare$}] The outer volatile envelope has insufficient mass and thus gravitational
pressure to significantly affect the equation-of-state of the inner layers.
\item[{\tiny$\blacksquare$}] The densest core permitted is that of iron i.e. heavy-element (e.g. uranium)
cores are forbidden.
\item[{\tiny$\blacksquare$}] We have accurate models for a planet's mass and radius given a
particular compositional mix.
\end{itemize}

We stress that what has been described thus far includes the possibility of a
light volatile envelope, and is not limited to some special case of two- or
three-layer conditions, as discussed earlier.

Under these assumptions, the limiting core radii fractions should be determinable, although
violating any of the assumptions listed above would invalidate our argument.
An obvious one is that the theoretical models used are invalid or
inaccurate, for example because their assumed equations of state are wrong. In
this work, we primarily use the \citet{zeng:2013} model but we point out that
should a user believe an alternative model to be superior, it is
straight-forward to reproduce the methods described in this paper using the
model of their preference. The actual existence of a boundary condition remains
true.

Another more serious flaw would be if the planetary body in question has
a significant mineral fraction based on some heavier element than iron, for
example a uranium core. Such a body could feasibly have a significantly smaller
core than that derived using our approach. If evidence for such cores emerges
in the coming years, then we advice against users employing the model described
in this work.

The remaining two assumptions, a differentiated, non-young planet and a light
volatile envelope, mean that young systems are not suitable and gas giants
would not be either. In general then, the model described is expected to be
valid for most planets smaller than mini-Neptunes.

\subsection{A parametric interpolative model for $\crfmin$}
\label{sub:interpolation}

For any combination of mass and radius, we need to be able to predict what
the corresponding CRF would be for a silicate-iron two-layer model, in
order to determine $\crfmin$. Inferring $\crfmax$ is far simpler and is
briefly explained later in Section~\ref{sub:crfmaxmethod}. In what
follows, we use the models of \citet{zeng:2013}, which are made available as a 
regular grid of theoretical points. Whilst we could simply perform a nearest
neighbor look-up, this is unsatisfactory since our precision will be limited to
the grid spacing and resulting posteriors would be rasterized to the same grid
resolution. Instead, we seek a means to perform an interpolation of the grid.

The first successful literature interpolation of the \citet{zeng:2013} models
comes from \citet{kipping:2013}, who found that for a specific fixed CRF,
each of the various two-layer models of \citet{zeng:2013} are very
well-approximated by a seventh-order polynomial of radius with respect to
the logarithm of mass, given by

\begin{equation}
\frac{\orob}{\re} = \sum_{i=0}^{7} a_i(CRF) \times \log \Big(\frac{\omob}{\me}\Big)^i
\label{equation2}
\end{equation}

Temperature does not feature in this expression, as \citet{zeng:2013}
find its effects on the density profile are secondary compared to pressure and
can be safely ignored.
Equation~\ref{equation2} is attractive since it is parametric,
linear (and thus can be trained using linear least squares) and extremely fast
to execute as an interpolative model once the coefficients have been assigned.
Accordingly, we are motivated to pursue a similar strategy in what follows.

\begin{figure}
\begin{center}
\includegraphics[width=8.4cm,angle=0,trim = 5 5 0 0,clip=true]{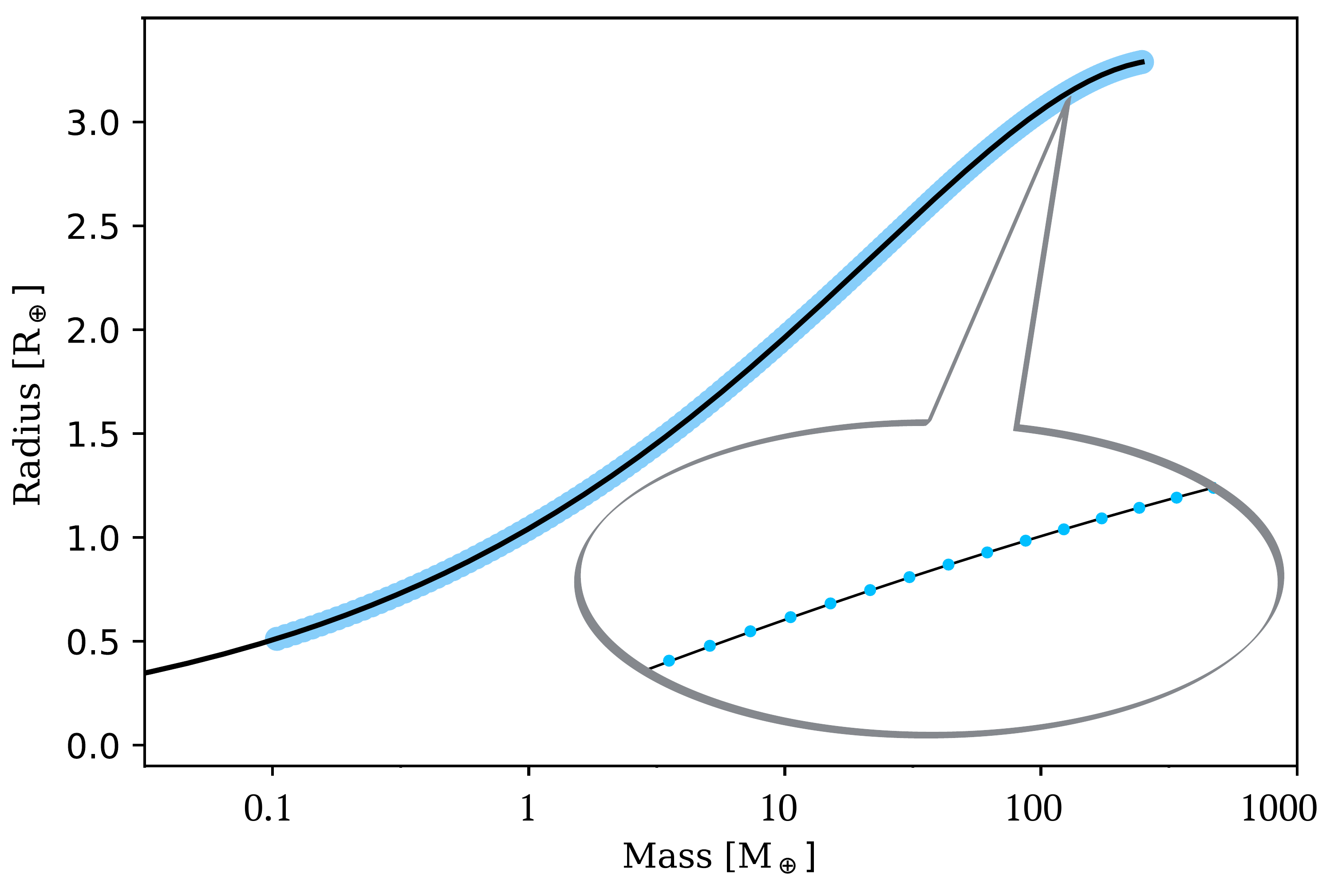}
\caption{
Theoretical mass-radius grid points of a pure silicate planet from
\citet{zeng:2013}, shown in blue. For comparison, we show our
$7^{\mathrm{th}}$ order polynomial fit. We only recommend this
interpolation for masses above 0.1\,$M_{\oplus}$.
}
\label{fig:7thorder}
\end{center}
\end{figure}

Consider first the mass-radius relation for a 100\% silicate planet. Plotting
the radius against the logarithm of the mass indeed reveals a series of points
that are well described by a seventh-order polynomial, as shown in 
Figure~\ref{fig:7thorder}. We found that the range for which this
interpolation works best is $M>0.1$\,$M_\oplus$ and thus we set this as a
truncation point during training. Note that any planet that lies beneath this
polynomial curve must have an iron core. 

\begin{figure}
\includegraphics[width=8.4cm]{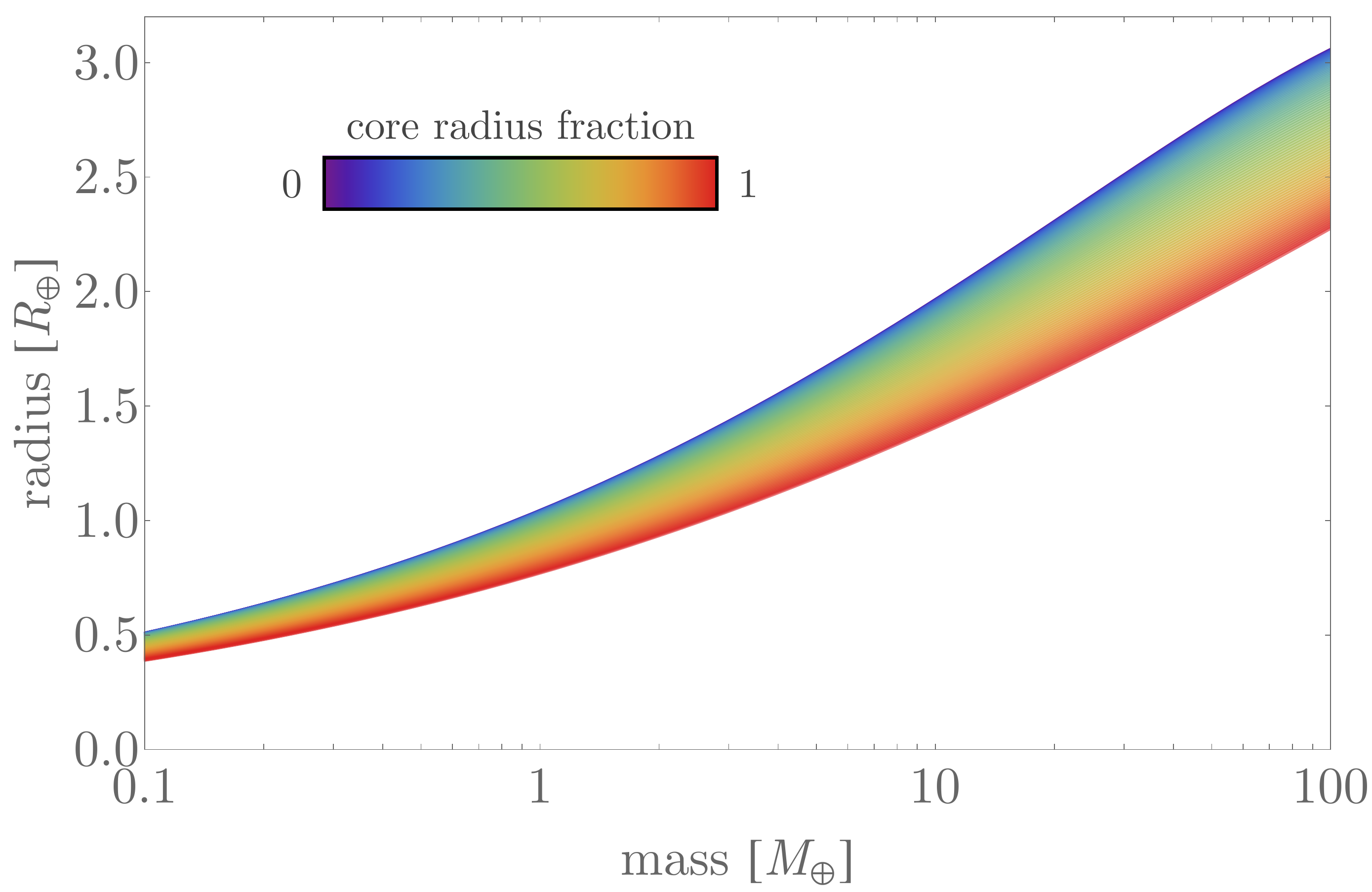}
\caption{
Interpolated theoretical mass-radius relations for a silicate-iron two-layer
solid planet for various core radius fractions (CRFs), based off
\citet{zeng:2013}. All interpolations for CRFs between 0 and 1 are
seventh-order polynomials. We are then motivated to describe the dependence of
the polynomials with respect to the CRF, by making the these coefficients
polynomial functions themselves.
}
\label{fig:rainbow}
\end{figure}

However, our goal is to not only determine whether or not a planet has an iron
core, but also quantify the minimum CRF. To accomplish this, we trained a suite
of seventh-order polynomials on models with varying CRFs assuming the two-layer
iron-silicate models of \citet{zeng:2013}. We varied the CRF from 0 to 1 in
0.025 steps and perform a linear least squares regression at each step.
To illustrate this, a gradient of interpolations of the mass-radius relation for
the CRFs between 0 and 1 are shown in Figure~\ref{fig:rainbow}. 

In order to have a general model, we are interested in parameterizing the CRF
variable to understand the relation of the polynomials shown in
Figure~\ref{fig:rainbow}. To do this, we allow the coefficients of the
polynomials to be polynomials themselves (but with respect to CRF rather than
logarithmic mass), such that

\begin{equation}
a_i(CRF) = \sum_{j=0}^{M_i} b_{i,j} \mathrm{CRF}^{j},
\label{equation3}
\end{equation}

where $M_i$ is the polynomial order of the $i^{\mathrm{th}}$ coefficient
and $b_{i,j}$ is the $j^{\mathrm{th}}$ ``sub-coefficient'' of the
the $i^{\mathrm{th}}$ coefficient.

\begin{figure*}
\includegraphics[width=\textwidth,width=17.5cm,trim = 0 7 0 0,clip=true]{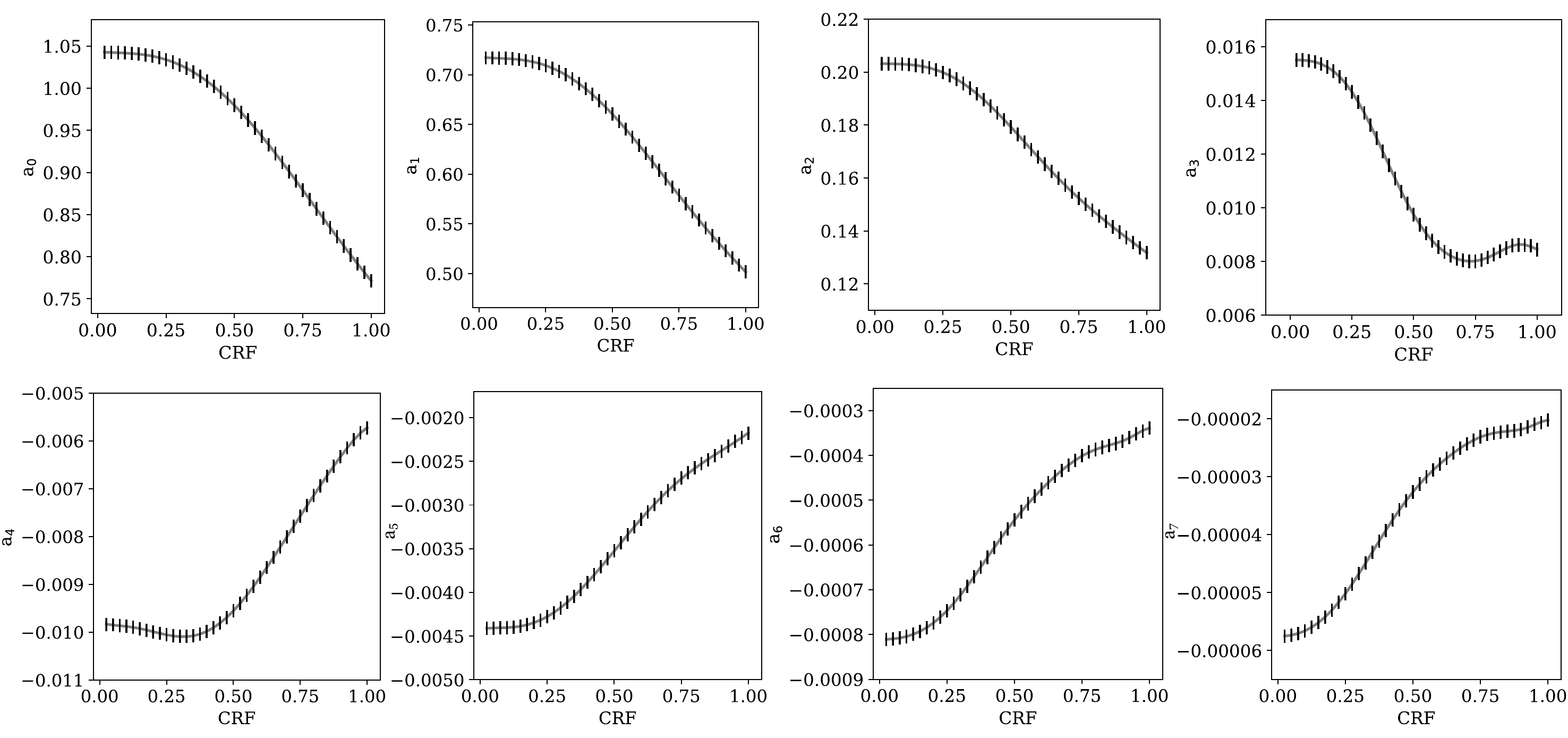}
\caption{
Interpolated functions for the coefficients of seventh-order polynomials. Given
a specific CRF, one can use this parameterization to solve for all eight
coefficients, to then use for a seventh-order polynomial. This seventh order
polynomial can then describe the mass-radius function corresponding to all
CRFs, not just the ones tabulated by \citet{zeng:2013}.
}
\label{fig:coeffs}
\end{figure*}

As an example, in the case of the $a_0$ coefficient, we first
graphed the coefficient for 40 steps as a function of the corresponding CRF
(see top-left panel of Figure~\ref{fig:coeffs}). It was immediately apparent
that the points followed a smooth function that could be stably approximated by
another polynomial. The polynomial-order was initially third-order and then
we stepped through until the polynomial appeared to go through almost
all of the data, ranging from fifth to tenth order. The same process was
repeated for the remaining eight coefficients. The resulting functions are
presented in Figure~\ref{fig:coeffs}, with the coefficients tabulated in
Table~\ref{tab:coefftab}.

\begin{table*}
\label{tab:coefftab}
\caption{
Coefficients for the terms in Equation~\ref{equation3}. These are implemented in our
package \hardcore\ available at \httplink.
}
%\centering % used for centering table
\begin{tabular}{lllllllll} % centered columns (7 columns)
\hline
j & $b_{0,j}$ & $b_{1,j}$ & $b_{2,j}$ & $b_{3,j}$ & $b_{4,j}$ & $b_{5,j}$ & $b_{6,j}$ & $b_{7,j}$ \\ [0.5ex] % inserts table
%heading
\hline
0 & 1.042859 & 0.717298 & 0.203201 & 0.015509 & -0.009827 & -0.004411 & -0.000811 & -0.000058 \\
1 & -0.022816 & -0.024719 & -0.001194 & -0.000338 & -0.001129 & 0.000238 & 0.000025 & 0.000002 \\
2 & 0.246925 & 0.292204 & 0.012756 & 0.017228 & 0.018847 & -0.004520 & 0.000727 & -0.000002 \\
3 & -1.525749 & -1.703841 & -0.076864 & -0.378486 & -0.175492 & 0.031420 & 0.021119 & 0.001470 \\
4 & 1.436881 & 1.827627 & -1.346393 & 2.642471 & 0.599478 & -0.033670 & -0.150980 & -0.010296 \\
5 & -0.406604 & -0.606841 & 3.407122 & -13.375869 & -0.886996 & -0.009464 &  0.712172 & 0.048466 \\
6 & 0 & 0 & -2.951612 & 39.839020 & 0.614844 & 0.028879 & -2.038051 & -0.142873 \\
7 & 0 & 0 & 0.884763 & -67.898807 & -0.165447 & -0.010646 & 3.449051 & 0.249493 \\
8 & 0 & 0 & 0 & 66.023432 & 0 & 0 & -3.387940 & -0.251336 \\
9 & 0 & 0 & 0 & -34.234323 & 0 & 0 & 1.788366 & 0.135208 \\
10 & 0 & 0 & 0 & 7.358617 & 0 & 0 & -0.392564 & -0.030096 \\ [1ex]
\hline %inserts single line
\label{tab:coefftab}
\end{tabular}
\end{table*}

To validate our model, we re-trained our model of all of the original training
data \citet{zeng:2013} but omitting a random single datum each time, serving as
a hold-out validation point. We then computed the relative difference between
the prediction for that point using the re-trained model, and the actual value.
Repeating for $10^4$ random hold-out points, we find that the mean error of our
model is 0.045\% and the maximum error is 0.24\%.

To assist the community, we make our model, which we dub \hardcore,
publicly available at \href{https://github.com/gsuissa/hardCORE}{this URL}. 

\subsection{A parametric model for $\crfmax$}
\label{sub:crfmaxmethod}

Determining $\crfmax$ is far more straight-forward than $\crfmin$. One may
simply take the 100\% iron mass-radius models, in our case from
\citet{zeng:2013}, and directly compute the expected radius of a pure iron
planet given an observed mass, $R_{\mathrm{iron}}(M_{\mathrm{obs}})$.

The maximum core radius fraction is then easily computed as
$\crfmax = R_{\mathrm{iron}}(M_{\mathrm{obs}})/R_{\mathrm{obs}}$. In practice,
this inversion is far simpler than $\crfmin$ since we need not interpolate
across intermediate mixtures of iron and silicate compositions, but rather
cam simply directly solve for $\crfmax$ if an analytic expression for
$R_{\mathrm{iron}}(M)$ is available. This could be estimated using our
interpolative model but is more directly accessible by simply fitting
Equation~\ref{equation2} to the \citet{zeng:2013} grid for the specific
points corresponding to a pure iron composition, which were previously
presented in the final column of Table~1 of \citet{kipping:2013}.

\section{Solving for the CRF Limits}
\label{sec:inverting}

\subsection{From forward- to inverse-modeling}

Thus far we have described a method to solve for $\crfmax$ but not for
$\crfmin$. The silicate-iron interpolative model is a forward model in which
we begin with knowledge of both the planet's mass and minimum core radius fraction and compute the
corresponding radius. In practice, however, we are interested in the inverse
model, where we wish to determine $\crfmin$ from the mass and radius.

The nested coefficient structure makes the problem non-linear with respect to
$\crfmin$, yet it is one dimensional and found to be unimodal in practice. For these
reasons, it is amenable to a large number of possible optimization algorithms,
but in what follows we adopted Newton's method, since we are able to directly
differentiate our functions thanks to their parametric nature.

Specifically, in our implementation we minimize the following cost
function, $J$, with respect to one degree of freedom, the $\crf$:

\begin{align}
J &= (R_{\mathrm{Fe-Si}}(\crf;M_{\mathrm{obs}}) - R_{\mathrm{obs}})^2,
\end{align}

where $M_{\mathrm{obs}}$ and $R_{\mathrm{obs}}$ are the observed mass
and radius of the planet, and $R_{\mathrm{Fe-Si}}(\crf;M)$ is the radius
of a two-layer iron + silicate model with core radius fraction $\crf$
and mass $M$. The latter function is determined using the smooth parametric
interpolation model described in Section~\ref{sub:interpolation}.
Our inversion algorithm, starts at an initial guess of $\crfmin=0.5$
and then iterates by computing

\begin{align}
\crf_{i+1} &= \crf_i - \frac{ J(\crf_i) }{ [\mathrm{d}J/\mathrm{d}\crf](\crf_i) }.
\end{align}

To improve speed and stability, we impose a check as to whether
$R_{\mathrm{obs}}$ is below that of a pure iron planet of mass
$M_{\mathrm{obs}}$, in which case we fix $\crfmin=1$, or if the radius
exceeds that of an pure silicate planet of mass $M_{\mathrm{obs}}$
(where again we use the interpolative model of \citealt{kipping:2013}),
in which case we fix $\crfmin=0$.

\subsection{The Earth as an example}
\label{sub:earth}

Let us use the Earth itself as an example of our method. We took a
1\,$M_{\oplus}$ and $1$\,$R_{\oplus}$ planet and used the methods
described earlier to solve for $\crfmin$ and $\crfmax$, giving
$\crfmin=0.43$ and $\crfmax=0.77$.

In reality, the Earth is not perfectly described by the \citet{zeng:2013}
model and the core in particular is only $\sim 80$\% iron, with nickel and
other heavy elements comprising the rest. The mantle-core boundary occurs
at a radius of 3480\,km relative to the Earth's mean radius of 6371\,km
\citep{dziewonski:1981}, meaning that its actual $\crf=0.55$. Accordingly, our CRF bounds correctly bracket the true solution, as expected.

We go further by treating these limits as being bounds on a prior distribution
for CRF. Adopting the least informative continuous distribution for a parameter constrained
by only two limits corresponds to a uniform distribution. Sampling from said
distribution yields a marginalized CRF of $\crfmarg = 0.600\pm0.098$, which is
again fully compatible with the true value.

To test our inversions in a probabilistic sense, we decided to
create a mock posterior distribution of an Earth-like planet where $M \sim 
\mathcal{N}[1.0 M_{\oplus},0.01 M_{\oplus}]$ and
$R \sim \mathcal{N}[1.0 R_{\oplus},0.01 R_{\oplus}]$ (we also apply a
truncation to the distributions at zero to prevent negative masses/radii).
This is clearly an optimistic assumption but a more detailed investigation of
sensitivity for different relative errors is tackled later in
Section~\ref{sub:sensitivity}. Generating $10^5$ samples, we inverted each
sample as described earlier to produce a posterior for
$\crfmin$ and $\crfmax$. Our experiment returns near-Gaussian like
distributions for both terms with a mean and standard deviation given by
$\crfmin = (0.43\pm0.04)$ and $\crfmax = (0.7716\pm0.0080)$. This
establishes that the inversions are stable against perturbations around
physical solutions.

As a brief aside, we argued earlier in Section~\ref{sub:BC} that the principle
of exploiting the boundary condition of theoretical models to infer $\crfmin$ does not explicitly require solid planets and works for mini-Neptunes
too. To demonstrate this point with a specific example, let us return to the earlier thought experiment of the Earth as a gaseous planet consisting of a solid iron core surrounded by a
light H/He envelope. We consider that the mass and radius of the planet remain
the same as the real Earth, and that the envelope significantly influences the
radius but has negligible mass. Using the 100\% iron-model of \citet{zeng:2013}
and the corresponding $7^{\mathrm{th}}$ order polynomial interpolation of
\citet{kipping:2013}, we estimate that a 1\,$M_{\oplus}$ iron core would have a
radius of 0.77\,$R_{\oplus}$ and therefore the remaining 0.23\,$R_{\oplus}$ is
given by the light, H/He envelope as depicted earlier in
Figure~\ref{fig:degeneracy}. Accordingly, such a body would have a core
radius fraction exceeding our inferred \textit{minimum} value of $\crf$ 
(which was $\crf>0.43$), which is expected and self-consistent with our
definition of $\crfmin$.

We highlight that the counter-example described above is highly unphysical
though; $1$\,$R_{\oplus}$ are not expected to retain significant volatile
envelopes in mature systems, both from a theoretical perspective
\citep{lopez:2014,owen:2017} and an observational one \citep{rogers:2015,
chen:2017,fulton:2017}. 

\subsection{Sensitivity analysis for an Earth}
\label{sub:sensitivity}

A basic and important question to ask is what kind of precisions on a planet's
mass and radius lead to meaningful constraints on $\crfmin$? In other words,
what is the correspondence we might expect between $\{ (\Delta M/M) ,
(\Delta R/R) \}$ and $(\Delta\crfmin/\crfmin)$? This is key for designing
future observational surveys, where primary science objectives may center
around inferring internal compositions. To investigate this, we repeated
the retrieval experiment described in Section~\ref{sub:earth}, but varied the
fractional error on mass and radius away from the fixed 1\% value previously
assumed.

In total, we generated $81^2=6561$ experiments, where for each one we
generated a new mock posterior of $10^5$ mass-radius samples, which was then
converted into a posterior of $\crfmin$ and $\crfmax$. For each
experiment, we record the standard deviation of the resulting posteriors as $\Delta\crfmin$ and $\Delta\crfmax$. The errors on the mass and radius
were independently varied with a fractional error given by $10^x$, where $x$
was varied across a regular grid from -4 to $0$ in 0.05
steps, giving 81 grid points in each dimension, and thus 6561
across both. In all experiments, the underlying mass and radius posteriors
are generated assuming a mean of $\mu_M=1$\,$M_{\oplus}$ and
$\mu_R = 1$\,$R_{\oplus}$.

\begin{figure*}
\begin{center}
\includegraphics[width=17.0cm]{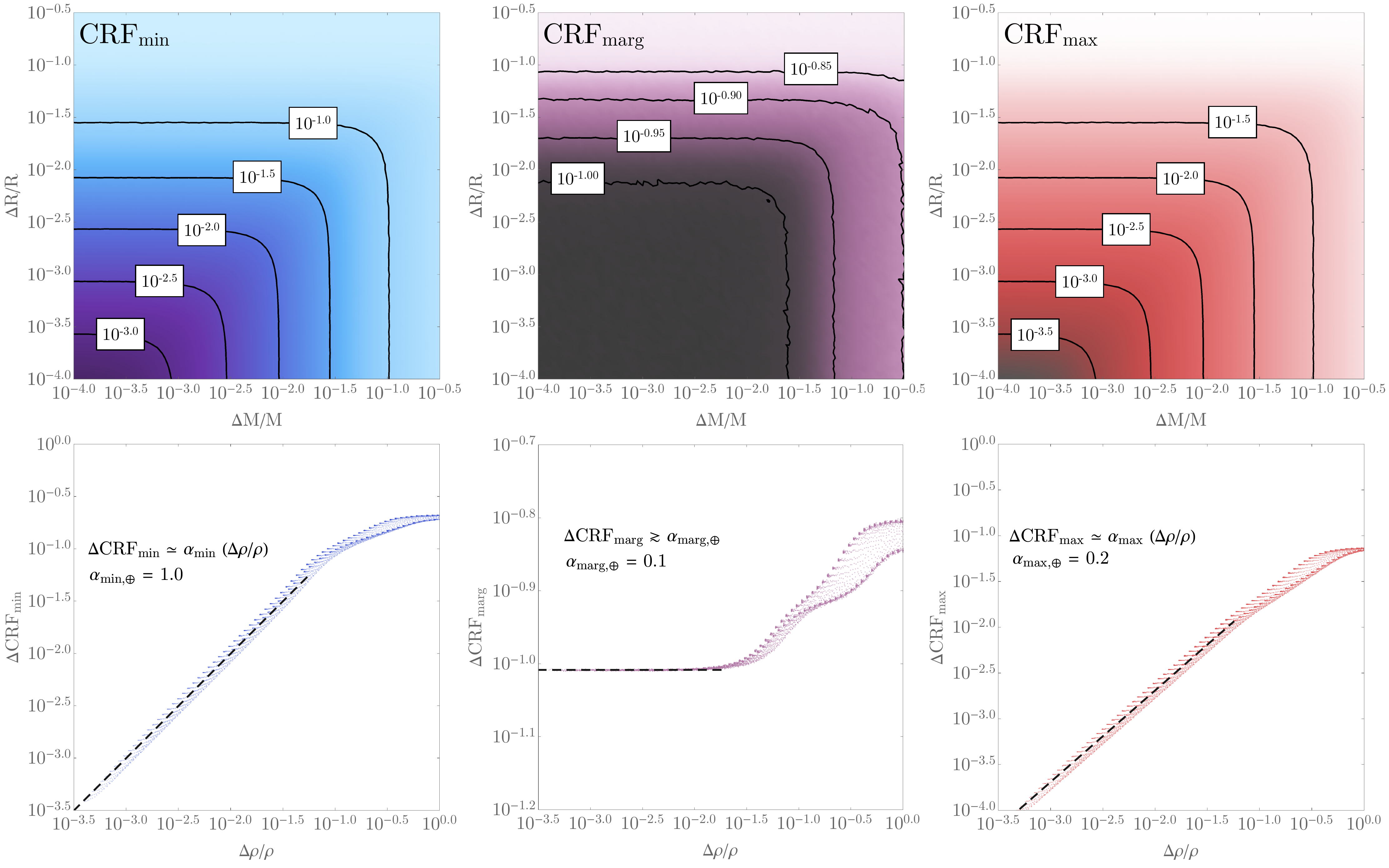}
\caption{
Upper: Contour plots depicting the a-posteriori standard deviation on the
minimum (left), marginalized (center) and maximum (right) CRF, as a function of
the fractional errors on mass and radius. Lower: Re-parameterization of the
above plots by combining the mass and radius axes into a single density term,
demonstrating a strong dependency in all cases.
}
\label{fig:sens}
\end{center}
\end{figure*}

Figure~\ref{fig:sens} displays the results of this effort for each combination
of mass and radius error. Given the shape of darker areas of the color plot,
one can see that radius is the dominant constraint, and that for the same
fractional error on mass and radius, it is the radius term which mostly strongly
constrains $\crfmin$. For example, we find that in order
to obtain a precision of 10\% on $\crfmin$, we require a measurement on the
mass better than 11\% and a measurement on the radius better than 3\%.

The ratio of these two numbers, close to three-to-one, led to us hypothesize
that density was the underlying driving term. This can be seen by calculating
error on density for independent mass and radius via

\begin{align}
\frac{\Delta\rho}{\rho} &= \frac{\Delta M}{M} + 3 \frac{\Delta R}{R}.
\end{align}

This is verified in the lower panels of Figure~\ref{fig:sens}, where we find
that although density doesn't perfectly capture the dependency, it describes
the vast majority of the variance. For precise densities ($\lesssim1$\%), the
dependency is strictly linear where we give the coefficients in the panels.
This linear dependency breaks down as the errors grow, likely as a result of
the truncated normals used to generate the masses and radii becoming
increasingly skewed and the finite support interval (zero to unity) of
the CRF itself causing a saturation effect.

The marginalized CRF behaves quite different to the other two. The upper-central
panel of Figure~\ref{fig:sens} alone looks fairly consistent with the previous,
just with inflated errors. This is to be expected by the very act of
marginalization. However, the bottom-central panel does not exhibit a simple
linear dependency, even at precise densities. In contrast, at precise densities
the marginalized CRF appears to saturate to ${\sim}10$\%. This implies that
no better than 10\% precision can ever be obtained on the $\crf$ using just
mass and radius alone.

\subsection{Generalized sensitivity analysis}
\label{sub:gensensitivity}

Thus far, we have assumed a $M=1$\,$M_{\oplus}$, $R=1$\,$R_{\oplus}$ planet.
In order to generalize the scalings found, we decided to vary these inputs and
repeat the entire process described above. We varied the mass from 1 to 10
Earth masses logarithmically and the CRF from 0.2 to 0.8 uniformly, exploring
over 1000 different realizations. For the $\crfmax$ term, we find that

\begin{align}
\Delta\crfmax &\simeq \alpha_{\mathrm{max}} \Bigg(\frac{\Delta\rho}{\rho}\Bigg)
\label{eqn:crfmaxsens}
\end{align}

provides an excellent fit across all simulations, where the best-fitting value
of the coefficient term ranged from $0.187<\alpha_{\mathrm{max}}<0.237$. The
relationship is sufficiently tight that it is reasonable to simply adopt
$\alpha_{\mathrm{max}}\simeq 0.2$ as a general rule of thumb. Repeating for
the minimum limit on the CRF we find that the function

\begin{align}
\Delta\crfmin &\simeq \alpha_{\mathrm{min}} \Bigg(\frac{\Delta\rho}{\rho}\Bigg)
\end{align}

again provides excellent fits, but now the coefficient term varies dramatically
from $0.6<\alpha_{\mathrm{min}}<4$ across all simulations, or logarithmically
a range of $-0.3<\log_{10}\alpha_{\mathrm{min}}<0.66$. The behaviour of
$\alpha_{\mathrm{max}}$ appears to display a peculiar and periodic dependency
with respect to the dependent variables, CRF and mass. We were unable to
identify a simple physically motivated relation after substituting for terms
such as density and surface gravity, but were able to capture the most of the
variance using an empirical approximate formula given by

\begin{align}
\log_{10}\alpha_{\mathrm{min}} \simeq -0.340985 + 0.0766358 \mathrm{exp} \beta_{\mathrm{min}},
\end{align}

where

\begin{align}
\beta_{\mathrm{min}} \simeq& \lfloor 4.934 (\crf - \tfrac{8}{3} \log_{10}M  - \tfrac{2}{3} \mathcal{R}[1.34 - 4 \log_{10}M] \rfloor \nonumber\\
\qquad& - 3.28933 \mathcal{R}[4 (\log_{10}M - 0.335)] - 4.934 \crf \nonumber\\
\qquad& + 13.1573 \log_{10}M.
\end{align}

where $\mathcal{R}$ is a rounding function.
We find the above function has a robust (via the median absolute deviation)
estimate of the RMS in the residuals of
$\Delta(\log_{10}\alpha_{\mathrm{min}}) = 0.058$.%, in other words it is
%accurate to within 13\%.

The marginal CRF plateau, which was 10\% in the case of the Earth,
also varies in a non-trivial manner, ranging from 2.8\% to 15.7\%
across our suite of simulations. Fortunately, the location of the plateau
appears to be directly related to $\alpha_{\mathrm{min}}$, and thus we can
use our earlier empirical function to predict this term too with

\begin{align}
\Delta\crfmarg &\simeq \alpha_{\mathrm{marg}}
\end{align}

where

\begin{align}
\log_{10}\alpha_{\mathrm{marg}} \simeq& -0.9819 - 0.00583 \mathrm{exp}(-13.6 \log_{10}\alpha_{\mathrm{min}}) \nonumber\\
\qquad& + 0.321 \log_{10}\alpha_{\mathrm{marg}}.
\end{align}

We briefly comment that this saturation-behavior can be understood as follows.
With imprecise data, the posteriors on the minimum and maximum limits will be
broad, and so sampling a point between them will yield an even broader
distribution. As the data become more precise, the posteriors distributions for
$\crfmin$ and $\crfmax$ converge towards sharp delta functions, but these two
limits have no reason or expectation to be on top of one another. Accordingly,
when we draw samples between them uniformly, we will still get a broad
distribution, albeit one less broad than that obtained when $\crfmin$ and
$\crfmax$ were also broad.

\section{Example Application}
\label{sec:applications}

To demonstrate the value of our prescription, we show here an example
application to a real exoplanetary system. As established in
Section~\ref{sub:sensitivity}, precise measurement errors on both
planetary mass and radius are required in order to infer $\crfmin$
at a meaningful level, certainly better than 10\% on both.

One of the best laboratories for precise measurements of planetary properties
comes from dynamically interacting systems, particularly those in near
mean motion resonances (MMRs) where transit timing variations (TTVs) may 
strongly constrain planetary mass \citep{holman:2005,agol:2005}. A good
example comes from the Kepler-36 system, where two planets gravitationally
perturb one another, enabling a measurement of both masses to better than
8\% precision \citep{carter:2012}. Coupled with precise planetary radii
at precisions better than 2.5\%, and the low-radius, high-density nature
of Kepler-36b in particular ($R_b=(1.486\pm0.035)$\,$R_{\oplus}$ and
$\rho_b=7.45_{-0.59}^{+0.74}$\,g\,cm$^{-3}$), Kepler-36 offers an excellent
test case for our technique.

To implement our code, we follow a similar procedure to that used in
Section~\ref{sub:earth}, except that we now use the real mass-radius joint
posterior distribution of Kepler-36b derived by \citet{carter:2012},
which features $10^4$ independent samples. Kepler-36c was not used
as the planet more likely resembles a gas giant than a terrestrial body,
as established using the MAH technique in \citet{kipping:2013}, who find
that a volatile envelope comprises at least 36\% of Kepler-36c's bulk radius.

\begin{figure}
\begin{center}
\includegraphics[width=8.4cm,angle=0]{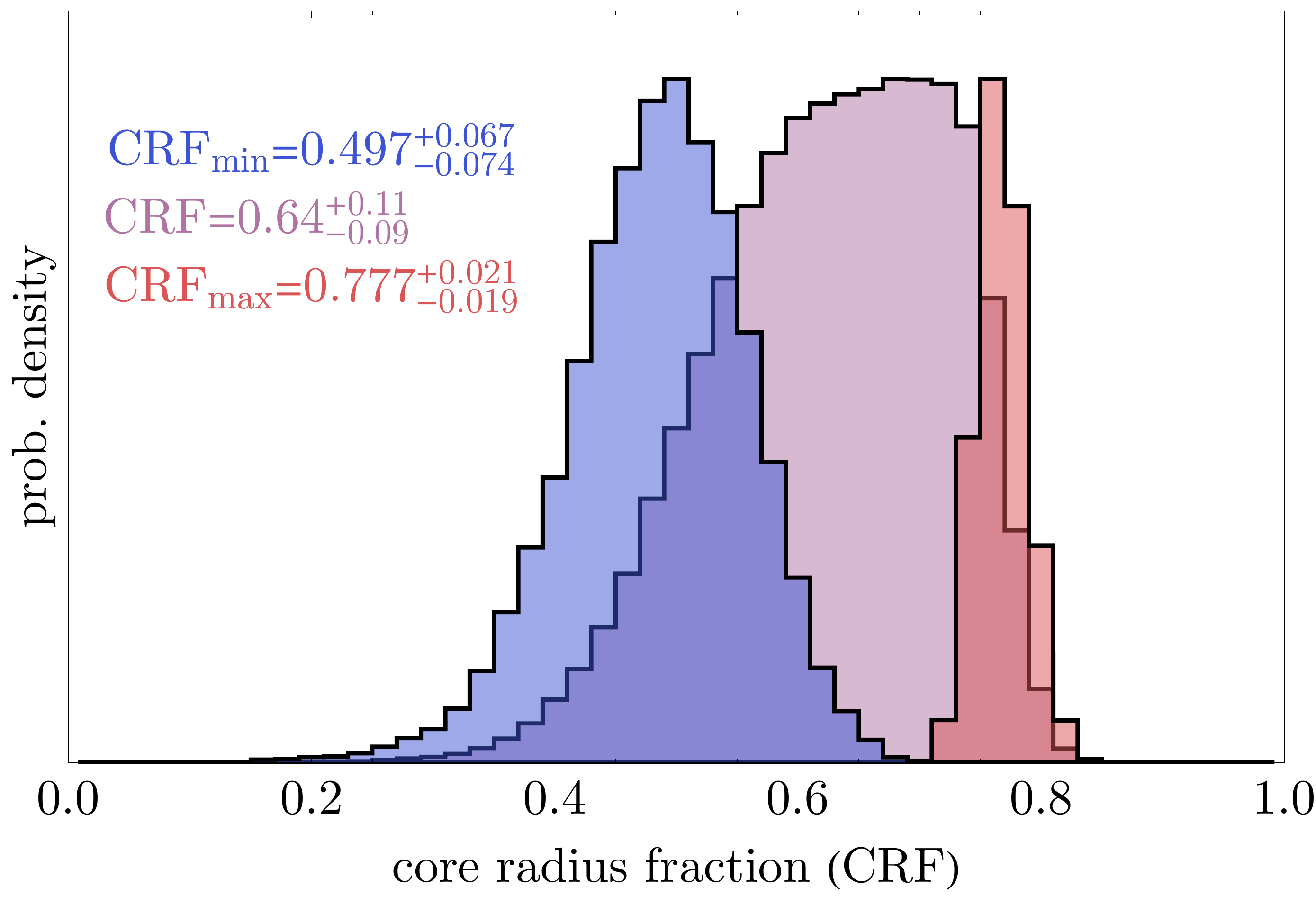}
\caption{
Posterior distribution of the minimum CRF (left),
maximum CRF (right) and marginalized CRF (center) for
Kepler-36b, based off the joint mass-radius posterior from 
\citet{carter:2012} and the model presented in this work.
Posterior heights normalized to be equivalent.
}
\label{fig:36b}
\end{center}
\end{figure}

The resulting posterior distributions on the $\crfmin$ and $\crfmax$ for
Kepler-36b are shown in Figure~\ref{fig:36b}. The $\crfmin$ posterior
indicates strong evidence for the presence of an iron core with
$\crfmin=0.497_{-0.074}^{+0.067}$. We obtain a tighter
constraint on $\crfmax$, as predicted by our earlier sensitivity analysis,
of $\crfmax=0.777\pm0.020$, which is exactly what would be predicted from
our scaling expression in Equation~\ref{eqn:crfmaxsens} for a 10\% error
on the planet's bulk density, which is indeed approximately the reported value.

By drawing random samples between $\crfmin$ and $\crfmax$ on a sample
by sample basis, we can construct the $\crfmarg$ posterior, revealing that
$\crfmarg=0.64\pm0.10$. Given that the fractional density error exceeds
${\sim}10^{-1.5}$, this precision does not lie on the sensitivity plateau
depicted in Figure~\ref{fig:sens}, and therefore we predict that the
precision on Kepler-36b's $\crfmarg$ could be improved. Since
\citet{carter:2012} only considered ten quarters, it should be possible
to considerably improve the uncertainties by re-analyzing the entire \Kepler\
time series.

For comparison, the Earth's $\crf$ is 0.55 but inverting $\crfmarg$ for a
synthetic Earth yields $\crfmarg = 0.60$, showing a slight bias in the
marginalization result (future work may be able to correct for this by
experimenting with different sampling schemes). Accordingly, our inference broadly agrees with the conclusions of previous authors studying
Kepler-36b's interior- that the planet appears to be compatible with having
an Earth-like interior \citep{dorn:2015,unterborn:2016,owen:2016}.
It is worth emphasizing this is by no means guaranteed and one
distinguishing feature of our approach is that it infers an Earth-like
composition rather than assuming it (e.g. see \citealt{lopez:2013}).

Recall that our approach is actually relatively simple compared to more
sophisticated approaches, such as leveraging stellar chemical proxies,
strictly two-layer modeling or hierarchical Bayesian methods. All we've
done is exploited basic boundary conditions in the problem. Despite
this, we have demonstrated its ability to reproduce the same inference
as other (often more complicated) models, giving us confidence that the
technique described here is a powerful and effective tool for the community to constrain the interior structure of solid planets.

\section{Discussion}
\label{sec:discussion}

In this work, we have presented a novel method for inferring the minimum
and maximum iron core radius fraction ($\crfmin$ \& $\crfmax$) of an
exoplanet using just it's measured mass and it's radius.
Building upon the earlier work of \citet{kipping:2013}, we exploit two boundary
conditions in the theoretical models describing solid exoplanet interiors. Our
method is valid under the assumptions of the specific underlying theoretical model
employed (we used the model of \citealt{zeng:2013} for example) and the
assumptions that the planet is differentiated, does not possess a high mass
volatile envelope (although light envelopes are fine) and that cores heavier than
iron are not permitted.

Although we used the theoretical models of \citet{zeng:2013} in this work,
the method can be easily adapted for whatever suite of theoretical models is 
preferred. Whilst $\crfmax$ is reasonably straight-forward to compute
(see Section~\ref{sub:crfmaxmethod}), $\crfmin$ has no simple solution. In order
to facilitate the inversion of the \citet{zeng:2013} models, we have developed
a parametric interpolation of the silicate-iron two-layer model from that work,
producing a function able to predict a planet's radius for any given mass and
core radius fraction. By iteratively inverting this expression with Newton's
method, we demonstrate that how the minimum CRF can be practically derived
too (details are provided in Section~\ref{sec:inverting}).

By drawing samples between the minimum and maximum limits, a marginalized
$\crf$ can be inferred. In this work, we have drawn samples assuming a uniform
distribution although we have highlighted that this may not be optimal, as it
appears to impose a slight positive bias in the results (see
Section~\ref{sec:applications}). As an applied example, we infer Kepler-36b's
internal core size to be $\crfmarg = 0.64\pm0.10$, compatible with, but
slightly larger than, that of the Earth (see Figure~\ref{fig:36b}).

By inverting our model on a suite of planets with differing measurement
precisions, we have produced a detailed sensitivity map for $\crfmin$,
$\crfmax$ and $\crfmarg$ (see Figure~\ref{fig:sens}). We find that
both terms have standard deviations proportional to the fractional density
uncertainty, although curiously $\crfmarg$ saturates in precision once the
density error hits ${\sim}$1-2\%, typically with a saturated standard deviation
ranging from 3-16\%. This indicates that mass and radius alone are unable to
improve upon our inference of the core meaningfully beyond this precision
limit, providing a clear goal post for observers interested in compositions.

Our approach is not the first, only nor likely the last, method proposed for
inferring planetary interiors. Generally speaking, any $N>2$ layer model
will be degenerate if inversion is attempted using just a mass and radius, and
our method circumvents this by exploiting boundary conditions in the problem.
However, it has been also proposed to exploit stellar metallicity constraints
from the parent star has a third datum to break the degeneracy
\citep{dorn:2015}, or simply adopt a two-layer model under certain reasonable
circumstances \citet{zeng:2016}. We encourage the community to view these
approaches as complementary, each operating under different assumptions but
ideally arriving at consistent inferences.

To aid the community it using our technique, we make our code fully public
(\hardcore\ available at \httplink) and it is extremely fast to execute and
perform inversions, providing an efficient tool for others to use in their
analyses.

This work highlights the challenging nature of inverting exoplanet interiors,
yet the potential to infer meaningful physical constraints. Precise masses
in particular will be crucial in future work in this area, with surveys
aiming to deliver sub m\,s$^{-1}$ radial velocities and long-term transit
timing variations being particularly valuable resources for this
enterprise.

\section*{Acknowledgments}

Thanks to members of the Cool Worlds Lab for helpful discussions in
preparing this manuscript.

%\appendix
%
%\input{lambda.tex}

\bsp
\label{lastpage}
\end{document}